\documentclass[12pt]{article}
\usepackage{epsfig}         

\begin{document}

\begin{center}
{\bf HADRONIC PRODUCTION OF DOUBLY CHARMED\\
BARYONS VIA CHARM EXCITATION IN PROTON}
\vspace{4mm}

D.A. G$\ddot u$nter and V.A.Saleev
\footnote{saleev@ssu.samara.ru}

{\it Samara State University, Samara, Russia}

\end{center}
\begin{abstract}
The production of baryons containing two charmed quarks $(\Xi
^*_{cc} \mbox{ or } \Xi_{cc})$  in hadronic interactions  at high
energies and large transverse momenta is considered. It is
supposed, that $\Xi_{cc}$-baryon is formed during a
non-perturbative fragmentation of the $(cc)$-diquark, which was
produced in the hard process of $c$-quark scattering from the
colliding protons: $c+c\to (cc) +g $. It is shown that such
mechanism enhances the expected doubly charmed baryon production
cross section on Tevatron and LHC colliders approximately 2 times
in contrast to predictions, obtained in the model of gluon - gluon
production of $(cc)$-diquarks in the leading order of perturbative
QCD.
\end{abstract}
%
%

\section{Introduction}
Doubly heavy baryons take the special place among baryons which
contain heavy quarks. The existence of two heavy quarks causes the
brightly expressed  quark - diquark structure of the $\Xi_{cc}$
baryon, in a wave function which one's the configuration with
compact heavy $(QQ)$-diquark dominates. Regularity in a spectrum
of mass of doubly heavy baryons appear in many respects to a
similar case of mesons containing one heavy quark \cite{1,2,3,4}.
Production mechanisms for  $(QQq)$-baryons and $(Q\bar q)$-mesons
also have common features. At the first stage  compact heavy
$(QQ)$-diquark  is formed, than it fragments in a final
$(QQq)$-baryon, picking up a light quark. The calculations of
production cross sections for doubly heavy baryons in $ep$- and
$pp$-interactions were made recently as in the model of a hard
fragmentation of a heavy quark in doubly heavy diquark
\cite{5,6,7} as within the framework of the model of precise
calculation of cross section of a gluon -- gluon fusion into
doubly heavy diquarks and two heavy antiquarks in the leading
order of the perturbation theory of QCD \cite{8,9,10}.

Mechanism of production of hadrons containing charmed quarks,
based on consideration of hard parton subprocesses with one
$c$-quark in an initial state, was discussed earlier in papers
\cite{11,12,13}. It was shown, that in the region of a large
transferred momentum $(Q^2
>> m_c^2$, where $m_c$ is charmed quark mass) the concept of a
charm excitation in a hadron does not contradict parton model and
allows to effectively  take into account the contribution of the
high orders of the perturbative QCD theory to the Born
approximation. However, there is open problem  of the "double
score", which is determined by the fact that the part of the Born
diagrams of birth of two heavy quarks in a gluon - gluon fusion
can be interpreted, as the diagrams with charm excitation in one
of initial protons. These diagrams give leading in $\alpha_s$
contribution in the $c$-quark perturbative, so-called point-like
structure function (SF) of a proton. As to the non-perturbative
contribution in $c-$quark SF of a proton \cite{14} it does not
depend from $Q^2 $ and becomes very small at $ Q^2 >> m_c^2 $.

For example, fig. 1 shows one of 36 Born diagrams, which have
order $\alpha_s^4$, describing production of the $(cc)$-diquark in
the gluon-gluon fusion subprocess. The
 experience in calculation of heavy quark production
cross sections in a gluon - gluon fusion demonstrates that the
contribution of the next order of the perturbative QCD in
$\alpha_s $ can be comparable with the contribution of the Born
diagrams. In the case of gluon -- gluon production  of the two
pairs of heavy quarks there will be more than three hundred
diagrams with additional gluon in the final state, which  have
order $\alpha_s^5$,  and their direct calculation is now
considered difficultly feasible.

\section{Subprocess $c+c\to (cc) +g$}
In this paper  the model of $(cc)$-diquark production in proton -
proton interactions, based on the mechanism of the charm
excitation in a proton is considered. It is supposed, that the
$(cc)$-diquark is formed during scattering of $c$-quarks from
colliding protons with radiation of a hard gluon, i.e. in the
parton subprocess:
\begin{equation}
c+c\to (cc) +g
\end{equation}
The Feynman diagrams of the parton subprocess (1) are shown in
fig. 2, where $q_1$ and $q_2 $ are 4-momenta of the initial
$c$-quarks, $k$ is 4-momentum of the final gluon, $p$ is
4-momentum of the diquark, which one is divided equally between
the final $c$-quarks. The doubly heavy diquark is considered as
bound state of two $c$-quarks in the antitriplet colour state and
in the vector spin state. If $i$ and $j$ are colour indexes of
initial quarks, and $m$ is colour index of a final diquark, the
amplitude of production of the $(cc)$-diquark $M_{ijm} (c+c\to
(cc) +g)$ is connected with the amplitude of production of two
$c$-quarks with 4-momenta $p_1=p_2 =\displaystyle{\frac{p}{2}}$ as
follows:
\begin{equation}
M_{ijm}(c+c\to(cc)+g,p)=K_0\displaystyle\frac{\varepsilon^{nmk}}{\sqrt{2}}
M_{ijnk} (c+c\to c+c+g, p_1=p_2 =\displaystyle{\frac{p}{2}}),
\end{equation}
where $K_0=\sqrt {\displaystyle {\frac{2}{m_{cc}}} }\Psi_{cc}(0)$,
$m_{cc}=2m_c $ is the diquark mass, $\Psi_{cc}(0)$ is the diquark
wave function in zero point,
$\displaystyle{\frac{\varepsilon^{nmk}}{\sqrt{2}}}$ is the colour
part of a diquark wave function. Considering spin degrees of
freedom of $c-$quarks and $(cc)$-diquark, we have following
conformity between amplitudes of birth of free quarks and a
diquark with fixed spin projections (without colour indexes and
common factor $K_0$):
\begin{eqnarray}
M (c+c\to (cc) +g, s_z = + 1) &\sim&
M (c+c\to c+c+g, s_{1z} = + 1/2, s_{2z} = + 1/2) \\
M (c+c\to (cc)+g,s_z=-1)& \sim& M
(c+c\to c+c+g, s_1z =-\frac{1}{2}, s_{2z} =-\frac{1}{2}) \\
M (c+c\to (cc) +g, s_z=0) &\sim& \frac{1}{\sqrt{2}}
\Biggl [M (c+c\to c+c+g,s_1z=+\frac{1}{2},s_{2z}=-\frac{1}{2})+\nonumber\\
&+&M(c+c\to c+c+g, s_{1z} =-\frac{1}{2}, s_{2z} = +\frac{1}{2})
\Biggr].
\end{eqnarray}
Because the wave function of the $(cc)$-diquark is antisymmetric
on colour index and  symmetric on remaining indexes, the
production of the scalar $(cc)$-diquark is forbidden, i.e.
\begin{equation}
M (c+c\to c+c+g, s_{1z} = +\frac{1}{2}, s_{2z}
=-\frac{1}{2})
-M (c+c\to c+c+g,s_{1z}=-\frac{1}{2},s_{2z}=+\frac{1}{2})=0.
\end{equation}
Amplitudes adequate to the diagrams in fig. 2, where the final
c-quarks are in the arbitrary spin states, are written out below,
without the colour factors and the common factor $K_0$:
\begin{eqnarray}
&&M_1=g_s^3\varepsilon_{\mu}(k)
\bar U(p_1)\gamma^{\mu}(\hat p_1+\hat k+m_c)
\gamma^{\nu}U(q_1)\bar U(p_2)\gamma_{\nu}U(q_2)
/{((p_1+k)^2-m_c^2)(p_2-q_2)^2}\\
&&M_2=g_s^3\varepsilon_{\mu}(k)
\bar U(p_1)\gamma^{\nu}U(q_1) \bar U(p_2)\gamma^{\mu}
(\hat p_2+\hat k+m_c)\gamma_{\nu}U(q_2)
/((p_2+k)^2-m_c^2)(q_1-p_1)^2\\
&& M_3=g_s^3\varepsilon_{\mu}(k)
\bar U(p_1)\gamma^{\nu}(\hat q_1-\hat k+m_c)
\gamma^{\mu}U(q_1)\bar U(p_2)\gamma_{\nu}U(q_2)
/((q_1-k)^2-m_c^2)(p_2-q_2)^2\\
&&M_4=g_s^3\varepsilon_{\mu}(k)
\bar U(p_1)\gamma^{\nu}U(q_1) \bar U(p_2)\gamma^{\nu}
(\hat q_2-\hat k+m_c)\gamma_{\mu}U(q_2)
/((q_2-k)^2-m_c^2)(q_1-p_1)^2\\
&&M_5=g_s^3\varepsilon_{\mu}(k)
\bar U(p_1)\gamma^{\nu}U(q_1) \bar U(p_2)\gamma^{\lambda}U(q_2)
G_{\lambda\mu\nu}(p_2-q_2,k,p_1-q_1)/{(q_1-p_1)^2(p_2-q_2)^2}
\end{eqnarray}
where $g_s=\sqrt{4\pi\alpha_s}$, $\alpha_s$ is strong coupling
constant, $G_{\lambda\mu\nu}(p,k,q)=(p-k)_\nu
g_{\lambda\mu}+(k-q)_\lambda g_{\nu\mu}+(q-p)_\mu g_{\nu\lambda}$.
Let's remark, that the amplitudes $M_6 - M_{10}$ are received by
replacement of the initial quarks momenta $q_1\leftrightarrow q_2$
in the amplitudes $M_1 - M_5$ ans are taken with a minus sign,
that allows for the antisymmetrization of the initial state of
two identical $c$-quarks. The corresponding colour factors are
presented by the following expressions:
\begin{eqnarray}
&&C_1=\frac{\varepsilon^{nmk}}{\sqrt
2}(T^c_{nl}T^b_{li})(T^b_{kj}), \quad
C_6=\frac{\varepsilon^{nmk}}{\sqrt
2}(T^c_{nl}T^b_{lj})(T^b_{ki}),\nonumber\\
&&C_2=\frac{\varepsilon^{nmk}}{\sqrt
2}(T^b_{ni})(T^c_{kl}T^b_{lj}), \quad
C_7=\frac{\varepsilon^{nmk}}{\sqrt
2}(T^c_{nj})(T^c_{kl}T^b_{li}),\nonumber\\
&&C_3=\frac{\varepsilon^{nmk}}{\sqrt
2}(T^b_{nl}T^c_{li})(T^b_{kj}), \quad
C_8=\frac{\varepsilon^{nmk}}{\sqrt
2}(T^b_{nl}T^c_{lj})(T^b_{ki}),\\
&&C_4=\frac{\varepsilon^{nmk}}{\sqrt
2}(T^b_{ni})(T^b_{kl}T^c_{lj}), \quad
C_8=\frac{\varepsilon^{nmk}}{\sqrt
2}(T^b_{nj})(T^b_{kl}T^c_{li}),\nonumber\\
&&C_5=\frac{i \varepsilon^{nmk}}{\sqrt
2}(T^b_{ni})(T^a_{kj})f^{bac}, \quad C_{10}=\frac{i
\varepsilon^{nmk}}{\sqrt 2}(T^b_{nj})(T^a_{ki})f^{bac}.\nonumber
\end{eqnarray}
Using known property of a completely antisymmetric tensor of the
third rank
$$\varepsilon_{n'mk'}\varepsilon^{nmk}=
\delta^n_{n'}\delta^k_{k'}- \delta^k_{n'}\delta^n_{k'},$$ it is
easy to find products of the colour factors $\displaystyle
C_{i,j}=\sum_{color} C_i C^*_j$, which ones are presented in the
Appendix A.

The method of the calculation of a production amplitude of a bound
nonrelativistic state of quarks in the fixed spin state is based
on a formalism of the projection operator \cite{15}. Using
properties of the charge conjugation matrix $C=i\gamma_2\gamma_0$,
we can link a scattering amplitude of a quark on a quark with a
scattering amplitude of an antiquark on a quark, for example:
\begin{eqnarray}
 &&M_1=g_s^3\varepsilon_\mu (k)
\bar U (p_1) \gamma^\mu (\hat p_1 +\hat k+m_c)
\gamma ^\nu U(q_1) \bar U(p_2) \gamma_\nu U (q_2) /
((p_1+k)^2-m_c^2)(p_2-q_2)^2=\nonumber\\
&&=g_s^3\varepsilon_\mu (k) \bar V (q_1) \gamma ^\nu (-\hat
p_1-\hat k+m_c) \gamma ^\mu V(p_1) \bar U (p_2) \gamma_\nu U(q_2)
/ ((p_1+k) ^2-m_c^2) (p_2-q_2) ^2.
\end{eqnarray}
As it may be shown, at $\displaystyle{p_1=p_2=\frac{p}{2}}$ one
has:
\begin{eqnarray}
&&V(p_1,s_{1z}=-\frac{1}{2})\bar U(p_2,s_{2z}=+\frac{1}{2}) \sim
\hat\varepsilon (p,s_z=+1)(\hat p+m_{cc}),\nonumber\\
&&V(p_1,s_{1z}=+\frac{1}{2})\bar U(p_2,s_{2z}=-\frac{1}{2}) \sim
\hat\varepsilon (p,s_z=-1)(\hat p+m_{cc}),\\
&&\frac{1}{\sqrt 2}\Bigl[V(p_1,s_{1z}=-\frac{1}{2})
\bar U(p_2,s_{2z}=+\frac{1}{2})+\nonumber\\
&&+V(p_1,s_{1z}=+\frac{1}{2})\bar U(p_2,s_{2z}=-\frac{1}{2})\Bigr]
\sim \hat\varepsilon (p,s_z=0)(\hat p+m_{cc}),\nonumber
\end{eqnarray}
where $\varepsilon^{\mu}(p)$  -- is polarization 4-vector of a
spin-1 particle. After following effective replacements
\begin{equation}
V (p_1) \bar U (p_2) \rightarrow \hat \varepsilon (p) (\hat
p+m_{cc}) \mbox{ and }K_0\rightarrow K,
\end{equation}
where $p_1=p_2=p/2$ and $\displaystyle
K=\frac{\Psi(0)}{2\sqrt{m_{cc}}}$, amplitudes $M_i$, with
corresponding colour factors $C_i$, describe production of the
$(cc)$-diquark with fixed polarization. The square of the module
of amplitude of $(cc)$-diquark production after average on spin
and colour degrees of freedom is given by the following
expression:
\begin{equation}
\overline{|{\cal M}|^2}= \displaystyle{\frac{1}{36}}K^2
\sum_{i,j=1}^{10}\sum_{spin} M_i(p_1,p_2)M_j^*(p_1,p_2) C_{i,j},
\end{equation}
where in  the amplitudes   $M_i$   we have put   $p_1=p_2=
\displaystyle{\frac{p}{2}}$. The summation on vector diquark
polarizations in the square of the amplitude of the process (1)
was done using the standard formula:
\begin{equation}
\sum_{spin}\varepsilon^{\mu}(p)
\varepsilon^{*\nu}(p)= -g^{\mu\nu} +
\displaystyle{\frac{p^{\mu}p^{\nu}}{m_{cc}^2}}.
\end{equation}
The calculation of the value
$F=\displaystyle{\sum_{i,j}^{10}\sum_{spin} M_i M_j^* C_{i,j}}$
have been executed using the package of an analytical calculations
FeynCalc \cite{16}. The answer is shown in the Appendix B, as a
function of standard Mandelstam variables $\hat s$ and $\hat t$.

\section{RESULTS OF CALCULATIONS}
In the parton model the cross section of a $(cc)$-diquark
production in $pp$-interactions is represented as follows:
\begin{eqnarray}
\displaystyle{\frac{d\sigma}{dp_\bot}}(pp\to (cc)+X)
&=&p_\bot\int^{y_{max}}_{y_{min}}dy
\int^{1}_{x_{1min}}dx_1C_p(x_1,Q^2)C_p(x_2,Q^2)
\times\nonumber\\
&& \times\displaystyle{\frac{\overline{|{\cal M}|^2}}
{16\pi(s(s-m_{cc}^2))^{1/2}} \times \frac{1}{x_1s-\sqrt{s}m_\bot
e^y}},
\end{eqnarray}
where
$$x_2=\displaystyle{\frac{x_1\sqrt{s}m_\bot e^{-y}-\frac{3}{2}m_{cc}^2}
{x_1s-\sqrt{s}m_\bot e^y}},$$
$$x_{1min}=\displaystyle{\frac{\sqrt{s}m_\bot e^y-\frac{3}{2}m_{cc}^2}
{s-\sqrt{s}m_\bot e^{-y}}},$$ $C_p (x,Q^2)$ is the $c$-quark
distribution function in a proton at
$Q^2=m_{\bot}^2=m_{cc}^2+p_{\bot}^2$, $p_{\bot}$ is the diquark
transverse momentum, $y$ is the rapidity of the diquark in c.m.f.
of colliding protons,
\begin{eqnarray}
&&\hat s=(q_1+q_2)^2=x_1x_2s+\displaystyle{\frac{m_{cc}^2}{2}},\nonumber\\
&&\hat t=(q_1-p)^2=\displaystyle{\frac{3}{2}m_{cc}^2}-x_1\sqrt{s}m_\bot
e^{-y},\\
&&\hat u=(q_2-p)^2=\displaystyle{\frac{3}{2}m_{cc}^2}-x_2\sqrt{s}
m_\bot e^y.\nonumber
\end{eqnarray}

It is supposed that spin-$\frac{1}{2}$ and spin-$\frac{3}{2}$
$\Xi_{cc}-$baryons relative yield is $1:2$ as it is predicted by
the simple counting rule for the spin states. The production cross
section of $\Xi_{cc}$-baryons plus $\Xi_{cc}^*-$baryons in our
approach is connected with the production cross section of
$(cc)$-diquark within the framework of a model of a
non-perturbative fragmentation as follows:
\begin{equation}
\displaystyle{\frac{d\sigma}{dp_\bot}(pp\to \Xi_{cc}+X)=
\int^1_0\frac{dz}{z}
\frac{d\sigma}{dp'_\bot}(pp\to (cc) X,p'_\bot=\frac{p_\bot}{z})
D_{(cc)\to \Xi_{cc}}(z,Q^2)},
\end{equation}
where $D_{(cc) \to \Xi_{cc}} (z, Q^2) $ is the phenomenological
function of a fragmentation, normalized approximately on unity, as
a total probability of transition $(cc)$-diquark in final doubly
charmed baryon. At $Q^2=m_{cc}^2$ the fragmentation function is
selected in the standard form \cite{17}:
\begin{equation}
D_{(cc)\to \Xi_{cc}}(z,Q^2)=\displaystyle{\frac{D_0}
{\displaystyle{z(m_{cc}^2-\frac{m_{\Xi}^2}{z}-\frac{m_q^2}{1-z})^2}}},
\end{equation}
where $m_{\Xi}=m_{cc}+m_q$ is the $\Xi_{cc}$-baryon mass, $m_q$ is
the light quark mass, $D_0 $ the is rate-fixing constant. The
fragmentation function for $Q^2>Q^2_0$ can be determined by the
solving the DGLAP evolution equation \cite{18}. Following to paper
\cite{10}, at numerical calculations we have used following values
of parameters: $m_{cc}=3.4 $ GeV, $ \alpha_s=0.2 $, $ | \Psi_{cc}
(0) | ^2=0.03 $ GeV$^3$, $m_q=0.3 $ GeV. For a $c-$quark
distribution function in a  proton $C_p (x, Q^2) $ the
parametrization CTEQ5 \cite{19} was used. In figures 3 and 4 at $
\sqrt s =1.8 $ TeV and $ \sqrt s =14 $ TeV, accordingly, the
curves show results of our calculations of $p_{\bot}$-spectra $(|y
| < 1)$  of $\Xi_{cc}$-baryons, the stars show results of the
calculations from paper \cite{10}, adequate to the contribution of
the gluon-gluon fusion production of $\Xi_{cc}$-baryons in a Born
approximation. Thus, our calculations demonstrate, that the
observed production cross section of $\Xi_{cc}$-baryons on
colliders Tevatron and LHC can be approximately 2 times more at
the expense of the contribution of the parton subprocess $c+c\to
(cc) +g $, than it was predicted earlier in the papers
\cite{8,10}.


The authors thank S.P.~Baranov, V.V.~Kiselev and A.K.~Likhoded for
useful discussions. The work is executed at support of the Program
"Universities of Russia -- Basic Researches" (Project 02.01.03).

\newpage
\centerline{\bf Appendix A}

\begin{center}
\begin{tabular}{lllll}
 $C_{1,1}=\frac{7}{9}$,&$C_{2,3}=\frac{10}{9}$,&$C_{3,6}=-\frac{10}{9}$,&$C_{4,10}=-2$,&$C_{6,10}=1 $\\
 $C_{1,2}=\frac{1}{9}$,&$C_{2,4}=-\frac{2}{9}$,&$C_{3,7}=\frac{2}{9}$,&$C_{5,5}=3$,&$C_{7,7}=\frac{7}{9} $\\
 $C_{1,3}=-\frac{2}{9}$,&$C_{2,5}=1$,&$C_{3,8}=\frac{8}{9}$,&$C_{5,6}=-1$,&$C_{7,8}=\frac{10}{9} $\\
 $C_{1,4}=\frac{10}{9}$,&$C_{2,6}=-\frac{7}{9}$,&$C_{3,9}=-\frac{16}{9}$,&$C_{5,7}=1$,&$C_{7,9}=-\frac{2}{9} $\\
 $C_{1,5}=-1$,&$C_{2,7}=-\frac{1}{9}$,&$C_{3,10}=2$,&$C_{5,8}=2$,&$C_{7,10}=1 $\\
 $C_{1,6}=-\frac{1}{9}$,&$C_{2,8}=\frac{2}{9}$,&$C_{4,4}=\frac{16}{9}$,&$C_{5,9}=-2$,&$C_{8,8}=\frac{16}{9} $\\
 $C_{1,7}=-\frac{7}{9}$,&$C_{2,9}=-\frac{10}{9}$,&$C_{4,5}=-2$,&$C_{5,10}=3$,&$C_{8,9}=-\frac{8}{9} $\\
 $C_{1,8}=-\frac{10}{9}$,&$C_{2,10}=1$,&$C_{4,6}=\frac{2}{9}$,&$C_{6,6}=\frac{7}{9}$,&$C_{8,10}=2 $\\
 $C_{1,9}=\frac{2}{9}$,&$C_{3,3}=\frac{16}{9}$,&$C_{4,7}=-\frac{10}{9}$,&$C_{6,7}=\frac{1}{9}$,&$C_{9,9}=\frac{16}{9} $\\
 $C_{1,10}=-1$,&$C_{3,4}=-\frac{8}{9}$,&$C_{4,8}=-\frac{16}{9}$,&$C_{6,8}=-\frac{2}{9}$,&$C_{9,10}=-2 $\\
$C_{2,2}=\frac{7}{9}$,&$C_{3,5}=2$,&$C_{4,9}=\frac{8}{9}$,&$C_{6,9}=\frac{10}{9}$,&$C_{10,10}=3$
\end{tabular}
\end{center}

\newpage
\centerline{\bf Appendix B}
\begin{equation}
F=-(4\pi\alpha_s)^3\frac{512F_N}{9F_D}
\end{equation}

\begin{eqnarray}
F_N&=&26361\,M^{18} - 6\,M^{16}\,\Bigl( 20513\,s + 67472\,t \Bigr)
+\nonumber\\
  &&+ 16\,M^{14}\,\Bigl( 14621\,s^2 + 100076\,s\,t - 86020\,t^2
\Bigr)-\nonumber\\
    &&-  16\,M^{12}\,\Bigl( 14873\,s^3 + 122408\,s^2\,t - 657280\,s\,t^2 -
         382560\,t^3 \Bigr)  +\nonumber\\
      &&+64\,M^{10}\,\Bigl( 2101\,s^4 - 658\,s^3\,t - 509652\,s^2\,t^2 -
         468736\,s\,t^3 - 170408\,t^4 \Bigr)  +\nonumber\\
      &&+65536\,s\,t^2\,{\Bigl( s + t \Bigr) }^2\,
       \Bigl( 9\,s^4 + 11\,s^3\,t + 13\,s^2\,t^2 + 4\,s\,t^3 + 2\,t^4 \
\Bigr)- \nonumber\\
&&- 256\,M^8\,\Bigl( 120\,s^5 - 8749\,s^4\,t - 201737\,s^3\,t^2
-\nonumber\\
        &&- 255896\,s^2\,t^3 - 149332\,s\,t^4 - 44640\,t^5 \Bigr)
-\nonumber\\
      &&-1024\,M^6\,\Bigl( 7\,s^6 + 2180\,s^5\,t + 44390\,s^4\,t^2
+\nonumber\\
        &&+ 74060\,s^3\,t^3 + 57876\,s^2\,t^4 + 28176\,s\,t^5 + 7184\,t^6 \
\Bigr)  -\nonumber\\
&&-16384\,M^2\,t\,\Bigl( 10\,s^7 + 353\,s^6\,t + 924\,s^5\,t^2
+\nonumber\\
         &&+1151\,s^4\,t^3 + 898\,s^3\,t^4 + 460\,s^2\,t^5 + 160\,s\,t^6 +
         28\,t^7 \Bigr)  +\nonumber\\
&&+4096\,M^4\,
       \Bigl( s^7 + 235\,s^6\,t + 5484\,s^5\,t^2 + 11610\,s^4\,t^3
+\nonumber\\
         &&+11609\,s^3\,t^4 + 7368\,s^2\,t^5 + 3056\,s\,t^6 + 672\,t^7 \Bigr)  \
\end{eqnarray}

\begin{equation}
F_D={\Bigl( M^2 - s \Bigr) }^2\,
    {\Bigl( M^2 - 4\,t \Bigr) }^4\,
    {\Bigl( 5\,M^2 - 4\,\Bigl( s + t \Bigr)
        \Bigr) ^4}
\end{equation}

\newpage

\newpage

\begin{figure}
\epsfig{figure=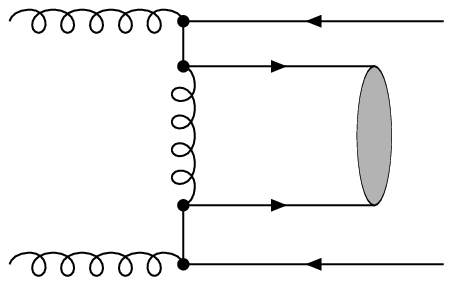,width=7cm}
 \caption{}

One of the Born diagrams
used for description subprocess $g+g\to (cc)+\bar c+\bar c$.
\end{figure}


\begin{figure}
\epsfig{figure=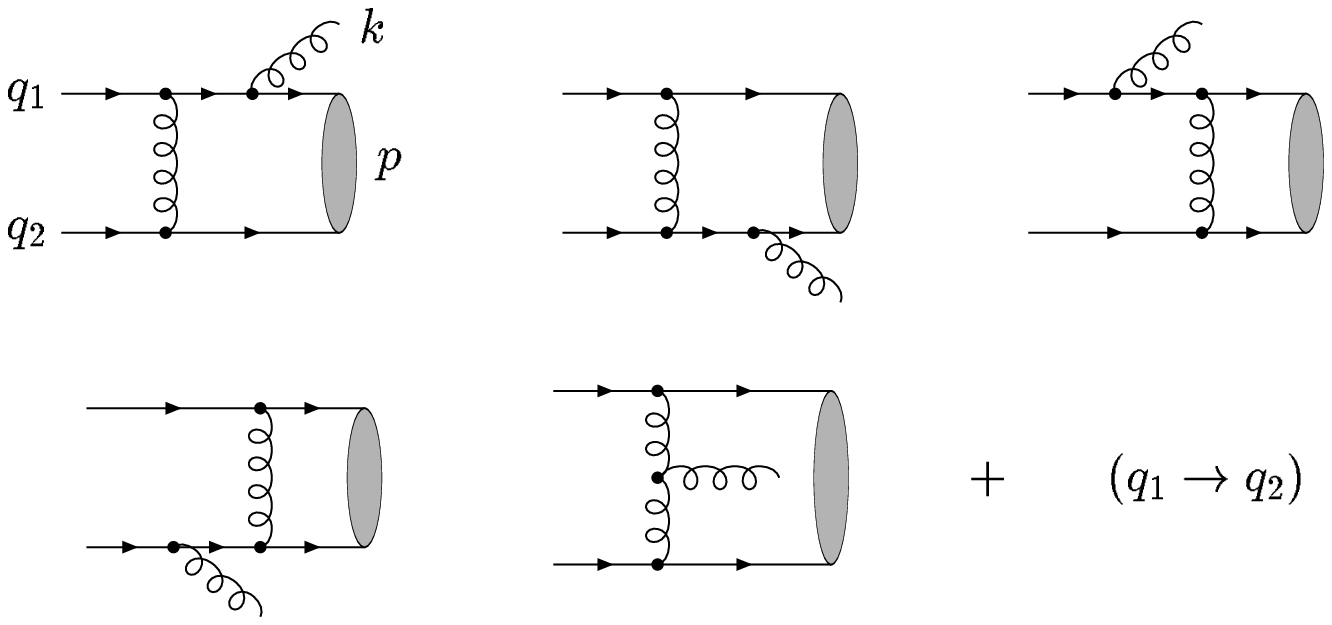,width=10cm,height=8cm} 
\caption{}

Diagrams
used for description subprocess $c+c\to (cc) +g$.
\end{figure}

\newpage

\begin{figure}
\epsfig{figure=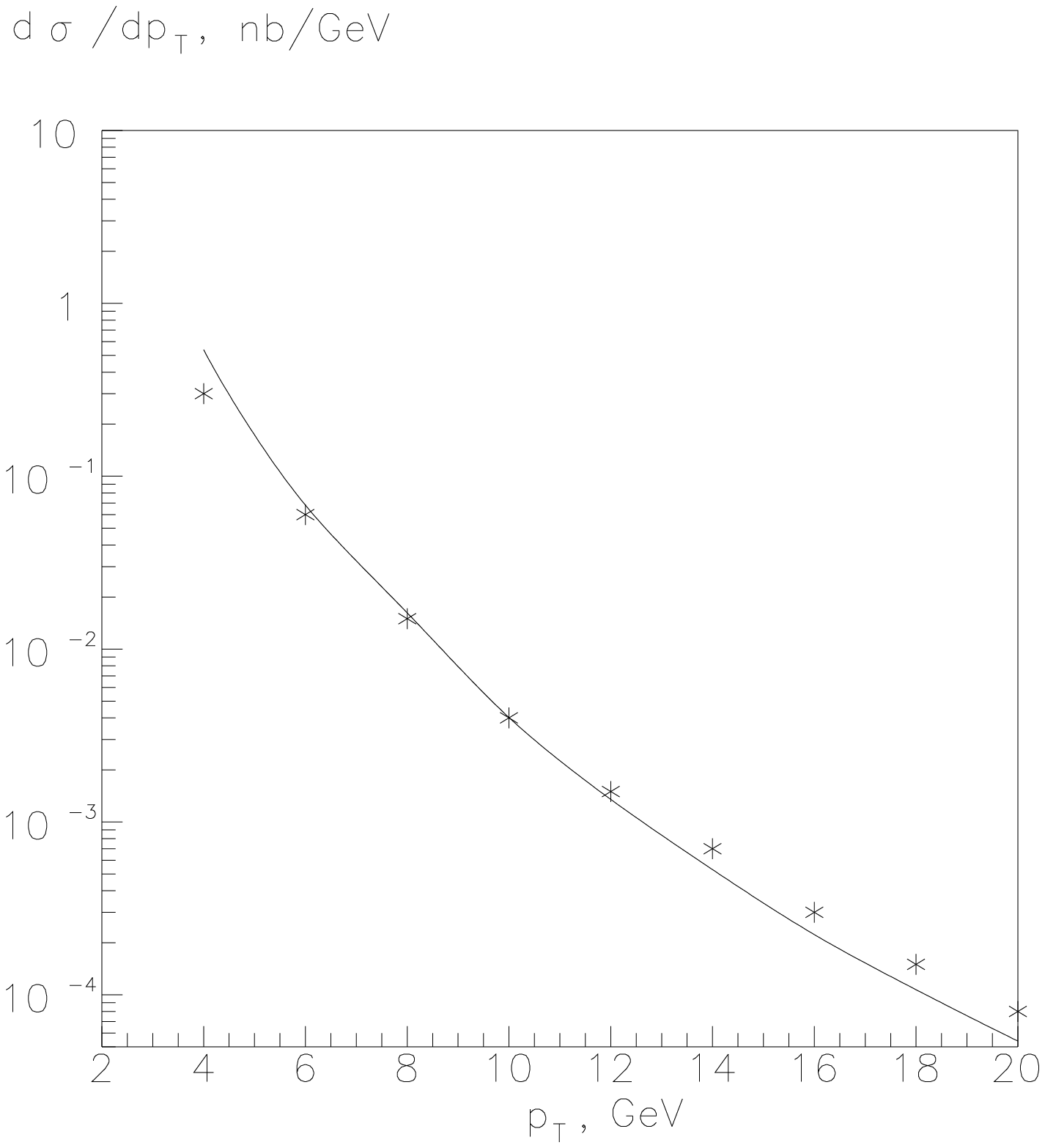,width=10cm} 
\caption{}

Cross section 
of $\Xi_{cc}$-baryon production at $\sqrt
s=1.8$ TeV and $|y|<1$. Stars (*) show the results of calculation
from paper [10], curve is our result obtained in the model of a
charm excitation in colliding protons.
\end{figure}

\newpage
\begin{figure}
\epsfig{figure=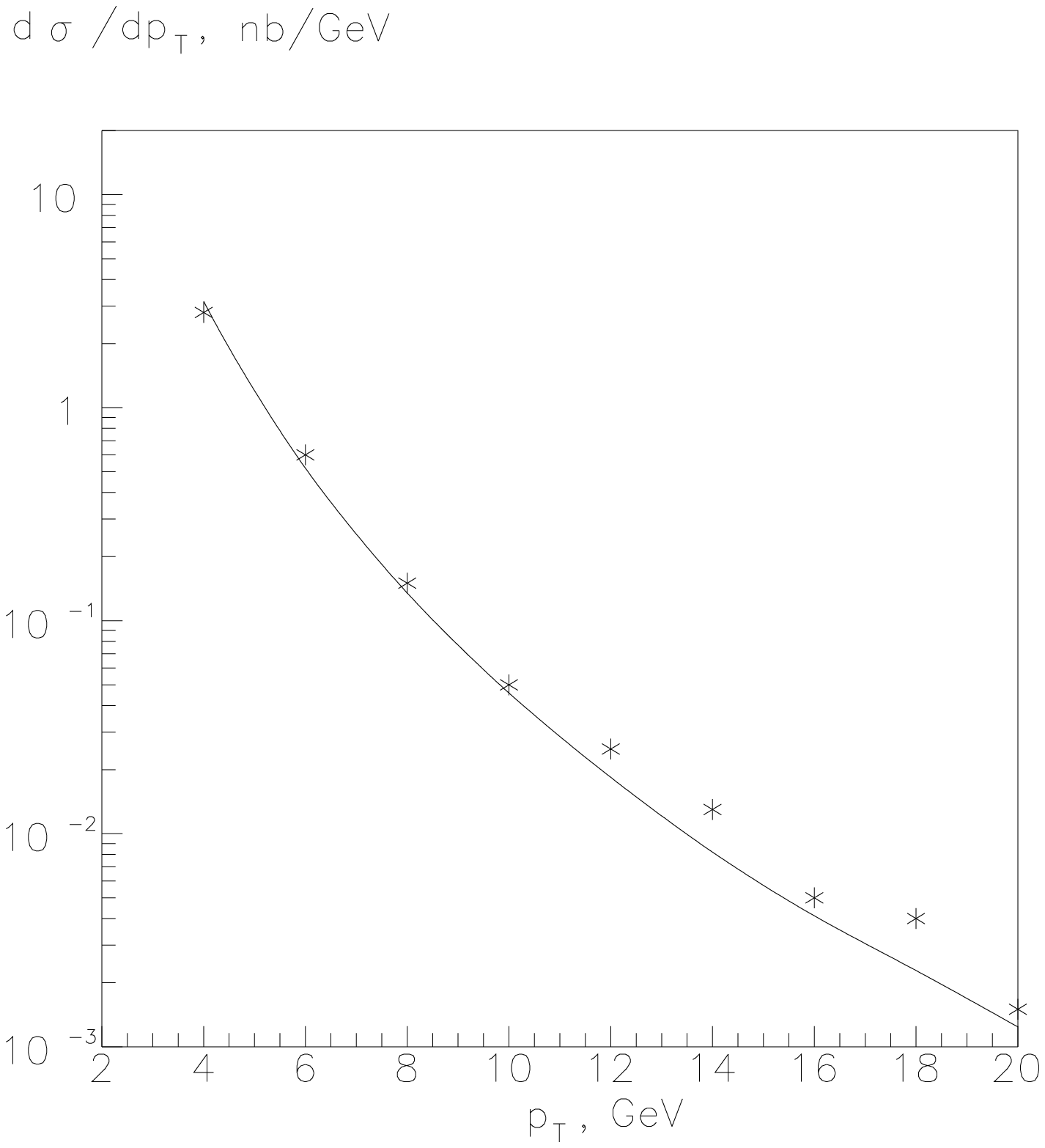,width=10cm} 
\caption{}
 Cross section 
of $\Xi_{cc}$-baryon production at $\sqrt
s=14$ TeV and $|y|<1$. Stars (*) show the results of calculation
from  paper [10], curve is our result obtained in the model of a
charm excitation in colliding protons.
\end{figure}

\end{document}